\documentclass[12pt,a4paper]{article}

\usepackage{latexsym}
\usepackage{amsmath}
\usepackage{amsfonts}
\usepackage{amssymb}

\newcommand{\be}{\begin{equation}}
\newcommand{\ee}{\end{equation}}
\newcommand{\bea}{\begin{eqnarray}}
\newcommand{\eea}{\end{eqnarray}}

\def\ad{{\mathrm{ad}}}                  %
\def\Ad{{\mathrm{Ad}}}                  %
\def\G{{\cal G}}                        %
\def\M{{\cal M}}                        %
\def\End{{\mathrm{End}}}                %
\def\cK{{\check {\cal K}}}              %
\def\cO{{\cal O}}                       %
\def\K{{\cal K}}                        %
\def\F{{\cal F}}                        %
\def\cR{{\cal R}}                       %
\begin{document}

\vspace*{0.5cm}
\begin{center}
{\Large \bf Spin Calogero models and dynamical
$r$-matrices\footnote{Based on talk by L.F. at  Symposium QTS-4, Varna (Bulgaria),
August 2005; to appear in the proceedings.}
}
\end{center}

\vspace{0.2cm}

\begin{center}
L. FEH\'ER${}^{a}$ and B.G. PUSZTAI${}^b$ \\

\bigskip

${}^a$Department of Theoretical Physics, MTA  KFKI RMKI\\
1525 Budapest 114, P.O.B. 49,  Hungary, and\\
Department of Theoretical Physics, University of Szeged\\
Tisza Lajos krt 84-86, H-6720 Szeged, Hungary\\
e-mail: lfeher@rmki.kfki.hu

\bigskip
${}^b$Centre de recherches math\'ematiques, Universit\'e de Montr\'eal\\
C.P. 6128, succ. centre ville, Montr\'eal, Qu\'ebec, Canada H3C 3J7, and\\
Department of Mathematics and Statistics, Concordia University\\
7141 Sherbrooke W., Montr\'eal, Qu\'ebec, Canada H4B 1R6\\
e-mail: pusztai@CRM.UMontreal.CA

\end{center}

\vspace{0.2cm}

\begin{abstract}
The main point of the construction of spin Calogero type classical integrable systems
based on dynamical $r$-matrices, developed by  L.-C. Li and  P. Xu,
is reviewed. It is shown that non-Abelian dynamical $r$-matrices
with variables in a reductive Lie algebra $\F$ and their Abelian counterparts with
variables in a  Cartan subalgebra of $\F$ lead essentially to the same models.

\end{abstract}

\newpage

\section{Introduction}
\setcounter{equation}{0}

Integrable systems of Calogero \cite{Cal}  (Sutherland \cite{Sut}, Moser \cite{Mos},
Olshanetsky-Perelomov \cite{OP}, Gibbons-Hermsen \cite{GH},
Ruijsenaars-Schneider \cite{RS} \dots) type
are related to many important areas of physics and mathematics.
The integrability of dynamical systems  is in general due to the
existence of conserved quantities that reflect some
(hidden) symmetries.
These are usually exhibited by constructing a Lax representation
for the equation of motion, and often also by deriving the system
of interest as a projection of a `free' system  which is integrable obviously.
(See \cite{Per} for a review.)
For Hamiltonian systems,
Liouville integrability is linked \cite{BV} with the
St Petersburg form of the Poisson brackets (PBs) of the Lax matrix, $L$, according to the
formula
\begin{equation}
\{ L_1, L_2\} = [ \rho, L_1] - [\rho_{21}, L_2 ],
\label{V1.1}\end{equation}
where $\rho$ is a $\G\otimes \G$ valued function on the phase space in general
if $L$ is $\G$-valued.
Here, $\G$ can be any Lie algebra and for simplicity we restrict ourselves
to spectral parameter independent cases.
(Note that $\rho_{21}=Y^a\otimes X_a$ if $\rho=X_a\otimes Y^a\in \G\otimes \G$,
and $L_1= L\otimes 1$, $L_2=1\otimes L$.)
Equation (\ref{V1.1}) guarantees
that the $\G$-invariant functions (eigenvalues)  of $L$ Poisson commute.
In the simplest cases, like for Toda systems, $\rho$ is a constant.
The $r$-matrices entering  (\ref{V1.1}) for the $A_n$ type
Calogero models  were found to be coordinate dependent \cite{AT,Skly,BS}.
They were re-derived in an inspiring way by Avan, Babelon and Billey \cite{ABB1, ABB2}
who  also related them to the classical dynamical Yang-Baxter equation (CDYBE)
that arose  from
conformal field theory \cite{BDF,Feld}.
Relying on the advance in the theory of the CDYBE  thanks to Etingof and
Varchenko \cite{EV}
and motivated also by the calculations in \cite{ABB1},
Li and Xu \cite{LiXu} proposed a method
to associate a spin Calogero model to
{\em any} dynamical $r$-matrix as defined in \cite{EV}.
This method was further developed by Li \cite{Li1,Li2,Li3,Li4} in
a rather abstract framework using Lie algebroids and groupoids.

In this report
we wish to contribute to the
`dynamical chapter' of the Yang-Baxter story  on integrability  by presenting
certain clarifications and applications of the  method invented by Li and Xu.
We focus on the spectral parameter independent version of the method
introduced in \cite{Li2}, and explain its essential point in a direct manner,
without any reference to Lie algebroids that feature in \cite{LiXu}-\cite{Li4}.
In principle, this method can be applied to
any dynamical  $r$-matrix defined on the dual space of
any Abelian or non-Abelian subalgebra of a Lie algebra.
However, we shall demonstrate  that the non-Abelian dynamical $r$-matrices
with variables belonging to a reductive Lie algebra, say $\F$,  and their
Abelian counterparts (Dirac reductions in the sense of \cite{FGP}) with
variables belonging to a  Cartan subalgebra of $\F$ lead essentially to the same models.
This `no go' result was mentioned in our recent paper \cite{FP},
where we studied spin Calogero type models built on {\em Abelian}
dynamical $r$-matrices. (It provided the reason for
considering there only Abelian $r$-matrices.)
We proved that the models based on $r$-matrices
with a certain non-degeneracy property are projections of the
natural geodesic system on a corresponding Lie group.
We shall briefly characterize these models in Section 4 at the end  of the present report,
referring to \cite{FP} for  details.

The most important new result of this paper is Proposition 2 in Section 3.
The content of Section 2 is not new, but it may be useful for
readers who want to learn about the essence of the method due to Li and Xu  keeping
the technicalities to a minimum.

\section{From dynamical $r$-matrices to integrable system}
\setcounter{equation}{0}

Consider a subalgebra $\K$ of a Lie algebra $\G$ and corresponding connected Lie
 groups $K$ and $G$.
Let
$\check \K^* \subset \K^*$ be an open subset invariant under the coadjoint action of $K$.
A {\em dynamical $r$-matrix} for $\K\subset \G$  is by definition \cite{EV} a
$K$-equivariant (smooth or holomorphic) map $r:\check \K^* \to \G\otimes \G$ satisfying
the CDYBE
\be
[r_{12}, r_{13}] + T_1^i \frac{\partial r_{23}}{\partial q^i}
+ \hbox{cycl. perm.} =0,
\label{2.1}\ee
and the additional condition that the symmetric part of $r$,
\be
r^s = \frac{1}{2} (r + r_{21}),
\label{2.2}\ee
is a $G$-invariant constant.
As usual
 $r_{23}= 1\otimes r$, $T_1^i = T^i\otimes 1\otimes 1$ etc, and
 $q^i\equiv \langle q, T^i\rangle$ are the components of $q\in \K^*$
 with respect to a basis  $T^i$ of $\K$.
Infinitesimally,  the $K$-equivariance property of $r$ reads
\be
[ T^i\otimes 1  + 1\otimes T^i , r(q) ]= f^{ji}_kq ^k
\frac{\partial r(q)}{ \partial q^j}
\quad\hbox{with}\quad
[T^i, T^j]= f^{ij}_kT^k.
\label{2.3}\ee
Important special cases are the quasi-triangular $r$-matrices with
symmetric part
$r^s=\frac{1}{2} T_\alpha\otimes T^\alpha$,
where $\G$ is self-dual with invariant scalar product $B_\G$,
$B_\G(T_\alpha, T^\beta) =\delta_\alpha^\beta$ for dual bases of $\G$, and
the triangular $r$-matrices characterized by $r^s=0$.
(The spectral parameter can be introduced in (\ref{2.1})
in the standard fashion.)

To construct integrable systems from dynamical $r$-matrices,
one starts with the phase space
\be
\M:= T^* \cK^* \times \G^* \simeq \check
 \K^* \times \K \times \G^*=\{ (q,p,\xi)\},
\label{2.4}\ee
and defines  the `quasi-Lax operator'
\be
L: \M \to \G, \qquad  L(q,p,\xi)= p - \cR(q) \xi,
\label{2.5}\ee
where $\cR(q) \in \End(\G^*,\G)$ corresponds to $r(q)\in \G\otimes \G$
so that
$X\otimes Y: \zeta\mapsto \langle \zeta,Y\rangle  X$ for any $X,Y\in \G$, $\zeta\in \G^*$.
Note that the map $L$ is equivariant with respect to the natural actions of the group
$K\subset G$ on $\M$ and on $\G$.
Introduce also the function
\be
\chi: \M \to \K^*, \qquad  \chi(q,p,\xi):= (\ad^\K_p)^*(q) +\xi_{\K^*},
\label{2.6}\ee
where
$\xi_{\K^*}\in \K^*$ is the
restriction of $\xi\in \G^*$ to $\K\subset \G$
and
$\langle (\ad^\K_p)^*(q), X \rangle = \langle q, [p,X]\rangle$ $\forall X\in \K$.
(We denote the pairing between any vector space and its dual by
$\langle\ ,\ \rangle$.)
By setting
\be
(\nabla_\chi r)(q,p,\xi) :=\frac{ d}{d t} r(q+ t \chi(q,p,\xi))\vert_{t=0},
\label{2.7}\ee
the fundamental result can be formulated as follows.

\medskip
\noindent
{\bf Proposition 1.}
{\em The quasi-Lax operator $L$ (\ref{2.5}) associated with any
dynamical $r$-matrix as defined above satisfies}
\be
\{ L_1, L_2\} = [ r, L_1 + L_2 ] - \nabla_\chi r.
\label{2.8}\ee
\medskip

\noindent
{\bf Proof.}
Let $\cR^a\in \End(\G^*,\G)$ (resp.~$\cR^s$) correspond
to the antisymmetric (resp.~symmetric) part of $r$.
Upon contraction with $1\otimes X\otimes Y$,
let us rewrite the CDYBE (\ref{2.1}) in the equivalent form
\be
E(\cR^a,X,Y)=- [\cR^s X, \cR^s Y],
\qquad
\forall X,Y\in \G^*,
\label{2.9}\ee
with  the $\G$-valued function $E(\cR^a,X,Y)$ on $\check \K^*$ given by
\be
E(\cR^a,X,Y)=[ \cR^a X, \cR^a Y]
-\cR^a \bigl(\ad_{\cR^a X}^\sharp Y -\ad_{\cR^a Y}^\sharp X\bigr)
+ \nabla_{Y_{\K^*}} \cR^a X
-\nabla_{X_{\K^*}} \cR^a Y +\langle X,(\nabla \cR^a) Y\rangle.
\label{2.10}\ee
Here $\ad^\sharp$ is the coadjoint representation of $\G$, $\ad^\sharp_T= - (\ad_T)^*$
($\forall T\in \G)$,
and
$\langle X,(\nabla \cR^a)Y\rangle=
T^i \frac{\partial \langle X,\cR^a Y\rangle}{\partial q^i} $.
To see that (\ref{2.1}) and (\ref{2.9})  are equivalent,
one must also use that $r^s$ is
a $\G$-invariant constant.
Now, for any $K$-equivariant $r^a$ for which $r^s$ is a $\G$-invariant constant,
one obtains from (\ref{2.5}) by an easy calculation
\be
\{ L_1, L_2\} - \left([ r, L_1 + L_2 ] - \nabla_\chi r\right)=
 \left( E(\cR^a, T_\alpha, T_\beta) + [\cR^s T_\alpha, \cR^s T_\beta]\right)
T^\alpha \otimes T^\beta
\label{2.11}\ee
with dual bases $T^\alpha\in \G$ and $T_\alpha\in \G^*$.
{\em Q.E.D.}
\medskip

It follows from Proposition 1 that
the $G$-invariant functions of $L$ yield a Poisson commuting family
{\em after} introducing the constraint $\chi=0$.
This is the basic idea for constructing integrable systems out of dynamical
$r$-matrices.
In coordinates,
the PBs on $\M= T^* \cK^* \times \G^*$ are
\be
\{ q^i, p_j\} = \delta^i_j\quad\hbox{and}\quad
\{ \xi^\alpha, \xi^\beta\} = f^{\alpha\beta}_\gamma \xi^\gamma,
\label{2.12}\ee
where $f^{\alpha\beta}_\gamma$  denote the structure constants of $\G$ (in a basis $T^\alpha$
extending the basis $T^i$ of $\K$, defining $\xi^\alpha = \langle \xi,T^\alpha\rangle$).
The constraints $\chi^i=0$ are {\em first class}, since
\be
\{ \chi^i, \chi^j\} = f^{ij}_k \chi^k.
\label{2.13}\ee
In fact, $\chi$ is nothing but the {\em momentum map} generating
the natural action
of the group $K$ on $\M$.
(The action of $K$ is induced by its coadjoint action on $\K^*$
and by composing the coadjoint
 action of $G$ on  $\G^*$ with the inclusion $K\subset G$.)
We perform Hamiltonian reduction by setting $\chi=0$. Thus we are
interested only in the gauge invariant ($K$-invariant) functions
on $\M^{\chi=0}$, i.e., in the reduced phase space $\M^{\chi=0}/K$.
In particular,
any $G$-invariant function $h$ on $\G$ yields a $K$-invariant function on $\M$ by $h\circ L$ as
 $L$ (\ref{2.5}) is a $K$-equivariant map.

For later purpose, note that one may also perform the Hamiltonian
reduction after restriction  to a symplectic leaf of $\M$, which  has the form
\be
T^* \check \K^* \times \cO,
\label{2.14}\ee
where $\cO\subset \G^*$ is a coadjoint orbit.
The reduction of the subspace (\ref{2.14}) of $\M$ leads to a union of symplectic
leaves in the full reduced phase space resulting from $\M$.

The constraint $\chi=0$ is universally applicable to remove the
derivative term of (\ref{2.8}), but in some examples non-zero constants
$\chi_0\in \K^*$ exist, too, for which $(\nabla_{\chi_0} r)(q)=0$ for
all $q\in \check \K^*$. Then the constraint $\chi=\chi_0$ can also be
used to obtain integrable systems. This occurs in particular
for the standard dynamical $r$-matrices on the Cartan
subalgebra $\K$ of $\G=u(n)$, for which $\chi_0$ can be taken as a
multiple of the unit matrix, after the usual identification
$\K^*\simeq \K$.
In our discussion we focus on the constraint $\chi=0$ for definiteness.

The (spectral parameter dependent  variant of the) basic formula (\ref{2.8}) first
appeared in \cite{ABB1}  for concrete examples of quasi-Lax
operators that were defined without referring to (\ref{2.5}).
The statement of Proposition 1 can be found in \cite{Li2},
and its spectral parameter dependent version for an Abelian $\K$
can be found in \cite{LiXu}.
Given (\ref{2.8}),
the idea to construct integrable systems by killing the derivative term arises
immediately and it occurs in all the references mentioned.
We thought it worthwhile to report the above elementary proof of Proposition 1,
because the results are presented in \cite{LiXu,Li2} in such an
abstract framework that may make it difficult to realize  how
simple the main point is.
Incidentally,
our proof clearly shows also that, in the presence of the equivariance
and invariance properties of $r^a$ and $r^s$, the PB relation (\ref{2.8}) for
the quasi-Lax operator (\ref{2.5}) does not only  follow from the CDYBE,
but is equivalent to it.
Formulae (\ref{2.5}), (\ref{2.8}) and the direct verification as above
work essentially in the same way for spectral parameter dependent $r$-matrices as well.

\section{Abelian versus non-Abelian dynamical $r$-matrices}
\setcounter{equation}{0}

In the first examples \cite{BDF,Feld}
the space of variables in the CDYBE  was a  Cartan subalgebra
of a simple Lie algebra. Later the concept was extended \cite{EV} to include
$r$-matrices defined on the duals of non-Abelian Lie algebras.
Such `non-Abelian $r$-matrices' came to light naturally in some
applications (see \cite{Feh,AM,EE}),
and Proposition 1 is valid in this general case.
At first sight, it appears a natural project to construct integrable systems from non-Abelian
$r$-matrices, and actually this had been one of our aims originally.
However, we found that such $r$-matrices do not give rise to new integrable systems
in addition to those that may be constructed using Abelian $r$-matrices, at
least if one considers $r$-matrices
on reductive Lie algebras of variables.
In the following we describe the derivation of this `no go' result.

Let $\G$ be a self-dual (also called quadratic) Lie algebra,
equipped with a non-degenerate, symmetric,
invariant bilinear form, $B_\G$.
Identify $\G\otimes \G$
with $\End(\G)$ in such a way that
$X\otimes Y: Z\mapsto B_\G(Y,Z) X$ for any $X,Y,Z\in \G$.
Consider a chain of subalgebras
\be
\K\subset\F\subseteq \G,
\label{3.1}\ee
where $\F$ is a {\em reductive} Lie algebra, $\K$ is a {\em Cartan} subalgebra of $\F$
and the restriction of $B_\G$ remains non-degenerate both on $\F$ and on $\K$.
Consider also a connected Lie group $G$ with Lie algebra $\G$ and connected subgroups
\be
K\subset F \subseteq G
\label{3.2}\ee
corresponding to the subalgebras (\ref{3.1}).
Let us assume that
\be
\cR_\F: \check \F \rightarrow \End(\G)
\label{3.3}\ee
is an $F$-equivariant map on a domain for which
\be
\ad_q \vert_{\K^\perp \cap \F}
\quad\hbox{is invertible}\quad \forall q\in
\check \K:= \K \cap \check \F,
\label{3.4}\ee
and $\check \F$ consists of orbits of $F$ through  $\check \K$, i.e.,
\be
\check \F = \{ \Ad_f q\,\vert\, f\in F,\, q\in \check \K\,\}.
\label{3.5}\ee
The properties expressed by the last two equations can always be arranged
by a restriction of the domain of any $F$-equivariant map $\cR_\F$.
We then define $\cR_\K: \check \K \rightarrow \End(\G)$ as follows:
\be
\cR_\K(q)(X)=\left\{
\begin{array}{cc}
\cR_\F(q)(X) &\mbox{if\,\, $X\in (\K+ \F^\perp)$}\\
\cR_\F(q)(X)+
\left(\mathrm{ad}_
q\vert_{\K^\perp\cap\F} \right)^{-1}(X) &\mbox{if\,\, $X\in \K^\perp\cap \F$}.
\end{array}\right.
\label{3.6}\ee
It is known that $\cR_\F$ is a solution of the CDYBE associated with $\F\subseteq \G$
{\em if and only if} $\cR_\K$ is a solution of the CDYBE associated with
$\K\subset \G$.
Of course, to view $\cR_\F$ and $\cR_\K$ as dynamical $r$-matrices,  one takes into account
the identifications $\F^*\simeq \F$ and $\K^*\simeq \K$ based on $B_\G$.
For reasons explained in \cite{FGP}, we call $\cR_\K$ the Dirac reduction of $\cR_\F$.
In the main examples \cite{AM} $\G$ is semi-simple with Killing form $B_\G$, and
$\F$ is the fixed point set of a (possibly trivial) automorphism of $\G$.
For a semi-simple Lie algebra $\G$, all non-Abelian $r$-matrices that are known to us
are related to corresponding Abelian $r$-matrices in the manner in (\ref{3.6}).

Now we show that the integrable systems that result by applying the construction
outlined in Section 2 to the non-Abelian $r$-matrix $\cR_\F$ and to its Abelian
counterpart
$\cR_\K$ are essentially (up to factoring by a discrete symmetry)  the same.
For the proof, it is convenient to proceed by first restricting the `spin' variable
$\xi\in \G^*$
to a coadjoint orbit $\cO\subset \G^*\simeq \G$, so that
the construction based on $\cR_\F$ starts with the
phase space
\be
\M_\F= T^* \check \F \times \cO =\check \F \times \F \times \cO = \{ (Q,P, \xi)\}.
\label{3.7}\ee
$\M_\F$ carries the symplectic form $\Omega_\F$,
\begin{equation}
\Omega_\F(Q,P,\xi)=
B_\G\left(dP \stackrel{\wedge}{,} dQ\right) + \omega_\cO(\xi),
\label{3.8}\end{equation}
where $\omega_\cO$ is the symplectic form of the orbit $\cO$,
and the quasi-Lax operator $L_\F$,
\be
L_\F(Q,P,\xi)  = P - \cR_\F(Q) \xi.
\label{3.9}\ee
The construction based on $\cR_\K$ works by reducing
$\M_\K= \check \K \times \K \times \cO = \{ (q,p, \xi)\}$,
which is
equipped with its symplectic form $\Omega_\K$,
\begin{equation}
\Omega_\K(q,p,\xi)=
B_\G\left(dp \stackrel{\wedge}{,} dq\right) + \omega_\cO(\xi),
\label{3.10}\end{equation}
and the quasi-Lax operator
$L_\K$ defined using $\cR_\K$.

Continuing with the Abelian case of $\cR_\K$, we decompose $\xi$ as
$\xi_\K + \xi_{\K^\perp}$ and introduce the first class constrained manifold
\be
\M_\K^0 = T^* \check \K \times \cO^0 =
\{ (q,p,\xi_{\K^\perp}) \,\vert\,
q\in \check \K,\, p\in \K,\, \xi_{\K^\perp}\in \cO\cap \K^\perp\,\},
\label{3.11}\ee
where $\chi_\K(q,p,\xi)=\xi_\K=0$.
The corresponding reduced phase space
\be
\M_\K^{red} = \M_\K^0/K
\label{3.12}\ee
is a (in general singular, stratified\footnote{To understand the fine structure
of the reduced phase spaces $\M_\K^{red}$ and $\M_\F^{red}$, one may wish to apply the
theory of singular symplectic reduction \cite{OR}.
This is directly applicable if $K$ and $F$ are compact Lie groups, but
actually our construction works without this assumption, too.})
symplectic manifold, whose symplectic structure is induced by the restriction (pull-back) of
$\Omega_\K$ to $\M_\K^0 \subset \M_\K$.
A commuting family of Hamiltonians on $\M_\K^{red}$ is obtained by the
application of the $G$-invariant
 functions on $\G$, $I(\G)\subset C^\infty(\G)$,
to the quasi-Lax operator
$L_\K$, since these Hamiltonians survive the reduction.
In fact,  $h\circ  L_\K$
 yields $\forall h\in I(\G)$
 a $K$-invariant function on $\M_\K^0$,
 on account of the $K$-equivariance of $L_\K$.

In the non-Abelian case of $\cR_\F$, we start by introducing the
first class constraints \be \chi_\F(Q,P,\xi) = [Q,P] + \xi_\F =0,
\label{3.13}\ee using the decomposition $\xi= \xi_\F+
\xi_{\F^\perp}$. This defines the constrained manifold
$\M_\F^0\subset \M_\F$, and we wish to compare $\M_\K^{red}$ to
$\M_\F^{red}= \M^0_\F/F$. The reduced (stratified) symplectic
structure and the commuting Hamiltonians on $\M_\F^{red}$ are
induced by the restrictions of $\Omega_\F$  and $h\circ L_\F$,
$h\in I(\G)$, to $\M_\F^0$. By the assumption (\ref{3.5}), every
$F$-orbit in $\M_\F^0$ intersects the submanifold
$\M_{\F,\K}^0\subset \M_\F^0$ defined by \be \M_{\F,\K}^0= \{
(q,P,\xi)\in \M_\F^0 \,\vert\, q\in \check \K\,\}, \label{3.14}\ee
i.e., any $Q\in \check \F$ can be conjugated into $\check \K$. Since
$\check\K$ consists of regular elements (\ref{3.4}), the `residual
gauge transformations' that preserve the `partial gauge fixing'
defined by $\M_{\F,\K}^0$ are given by the normalizer subgroup \be
N_F(\K):=\{\, f\in F\,\vert\, \Ad_f \kappa \in \K\quad \forall
\kappa\in \K\,\} \ee of $\K$ inside $F$. In other words, an
arbitrarily fixed element of $\M^0_{\F,\K}$ is mapped to
$\M^0_{\F,\K}$ precisely by those $f\in F$ that lie in $N_F(\K)$.
Note that $K\subset N_F(\K)$ is a normal subgroup, and \be W:=
N_F(\K)/K \label{3.15}\ee is a discrete group since the Lie algebra
of $N_F(\K)$ equals $\K$. These observations imply the second and
third equalities in \be \M_\F^{red}=\M_\F^0/F= \M_{\F,\K}^0/N_F(\K)
= (\M^0_{\F,\K}/K)/W. \label{3.16}\ee

In order to compare with the Abelian case, notice that on $\M^0_{\F,\K}$
the constraint (\ref{3.13})
 is uniquely solved as
\be \xi_\K=0, \qquad P=P(q,p,\xi)= p - \left(\ad_q \vert_{\K^\perp
\cap \F}\right)^{-1} \xi_{\K^\perp \cap \F} \quad\hbox{with}\quad
p\in \K, \label{3.17}\ee where we use the decomposition $\xi=\xi_\K
+ \xi_{\K^\perp\cap \F} + \xi_{\F^\perp}$. We see from this that the
map \be m: \M_\K^0 \to \M^0_{\F,\K}, \qquad m: (q,p,\xi)\mapsto (q,
P(q,p,\xi),\xi) \label{3.18}\ee is a $K$-equivariant
diffeomorphisms, and it is also easy to check that this map converts
the restrictions of the relevant symplectic forms and quasi-Lax
operators into each other: \be m^* \left(\Omega_\F\vert \M^0_{\F,\K}
\right) =\Omega_\K\vert \M^0_{\K} \quad\hbox{and}\quad m^*
\left(L_\F\vert \M^0_{\F,\K} \right) =L_\K\vert \M^0_{\K}.
\label{3.19}\ee Here, $\Omega_\K\vert \M_\K^0$ denotes $\imath^*
\Omega_\K$ with the natural map $\imath: \M_\K^0 \to \M_\K$ and,
similarly, $\Omega_\F\vert \M_{\F,\K}^0$ is the pull-back of
$\Omega_\F$. Since $m$ is $K$-equivariant, it induces a one-to-one
map, $\bar m: \M^{red}_\K \to \M_{\F,\K}^0/K$, whereby we can
identify these spaces of $K$-orbits. Because of (\ref{3.19}), this
identification converts the Poisson structures and commuting
Hamiltonians carried by these spaces into each other. By combining
the map $\bar m$ with the last equality in (\ref{3.16}), we arrive
at  the following conclusion.

\medskip
\noindent
{\bf Proposition 2.}
{\em Under the foregoing assumptions (in particular, choosing the domains
of $\cR_\F$ and $\cR_\K$ according to
(\ref{3.4})-(\ref{3.5})),
the Hamiltonian systems associated (using $\forall h\in I(\G)$)
with
the non-Abelian $r$-matrix $\cR_\F$ by the construction outlined in Section 2
are identical to the systems associated with
the Abelian $r$-matrix $\cR_\K$ (\ref{3.6}) up to the discrete symmetry given by
the group $W$ (\ref{3.15}).
That is the corresponding phase spaces are related as
\be
\M_{\F}^{red}= \M_\K^{red}/W.
\label{3.20}\ee
}

\medskip
\noindent
{\em Remark 1.}
Suppose that the semi-simple factor of $\F$ is compact, and notice that
$W$ is then the Weyl group of $\K\subset \F$.
Thus the space of $W$-orbits  $\M_\K^{red}/W$ can be realized simply by
restricting the variable $q$ to a fundamental domain of $W$ in $\check \K$, i.e.,
to an open Weyl chamber.
This means that the system on $\M_\F^{red}$ associated with $\cR_\F$
also arises by performing the
construction of Section 2 using $\cR_\K$ restricted to a Weyl chamber.
This strengthens our claim that the systems associated with the non-Abelian
and Abelian $r$-matrices are  `essentially' the same.

\medskip
\noindent
{\em Remark 2.}
A non-compact reductive Lie algebra
 $\F$ possesses non-conjugate Cartan subalgebras, say $\K_a$ for $a=1,\ldots, N>1$,
in general. If  $\cR_\F$ is defined on a {\em dense} open subset of $\F$
(which can be achieved for any $r$-matrix on $\F\subseteq \G$),
then the associated reduced phase space $\M^{red}_\F$ contains the
reduced spaces $\M_{\K_a}^{red}$ associated with the non-conjugate Cartan
subalgebras as disjoint open subsets.

\section{On  the resulting family of spin Calogero models}
\setcounter{equation}{0}

We  explain below that the dynamical $r$-matrix method
presented in Section 2 leads to a large family of
integrable systems of spin Calogero type.
The Hamiltonians of these systems are induced by the
quadratic form of an invariant scalar product using
Abelian dynamical $r$-matrices.

Let us take an Abelian,  self-dual subalgebra $\K$  of a
self-dual Lie algebra $\G$ and suppose that
\be
\cR: \check \K \to \End(\G)
\label{4.1}\ee
is a dynamical $r$-matrix for $\K\subset \G$,
where we made the identifications $\G\simeq \G^*$ and $\K\simeq \K^*$.
Suppose also that the operator $\cR(q)$  ($q\in \check \K$) is compatible with
the decomposition
\be
\G= \K + \K^\perp.
\label{4.2}\ee
This compatibility condition holds for all examples we are aware of.
The simplest Hamiltonian of interest
on the phase space $\M = \check \K \times \K \times \G$
is
\be
H(q,p,\xi) = \frac{1}{2} B_\G(L(q,p,\xi), L(q,p,\xi)),
\label{4.3}\ee
which corresponds to the
quadratic Casimir associated with the invariant scalar product $B_\G$.
Upon imposing the first class constraint
$\xi_\K=0$ on $\xi=\xi_\K + \xi_{\K^\perp}$ and recalling (\ref{2.5}),
the Hamiltonian takes the following form:
\be
H(q,p,\xi)= \frac{1}{2} B_\G(p,p) + \frac{1}{2} B_\G(\cR(q) \xi_{\K^\perp},
\cR(q)\xi_{\K^\perp}).
\label{4.4}\ee
If $\cR(q)$ depends on $q$ through
rational or
trigonometric (hyperbolic)  functions of its components,
which holds in all known examples,
then (\ref{4.4}) yields a Hamiltonian
of spin Calogero  type.
The first term of (\ref{4.4}) represents kinetic energy and the second
term is a rational or trigonometric (or hyperbolic) potential containing
the `spin' degrees of freedom as well.
The restriction of $B_\G$ to $\K$ must be positive or negative
definite for this interpretation to be valid in the strict sense.

Under the constraint $\xi_\K=0$, the evolution equation generated by $H$ (\ref{4.3})
implies
\be
\dot{L} = [\cR(q)L, L].
\label{4.5}\ee
Together with $\dot{q}=p$, the Lax equation (\ref{4.5}) is actually equivalent to the
constrained  Hamiltonian equation of motion if  $\cR(q)$
maps $\K^\perp$ to $\K^\perp$ in an {\em invertible} manner.
Indeed,
for such {\em non-degenerate} $r$-matrices $\dot{p}$ and
$\dot \xi_{\K^\perp}$ can be
recovered from the decomposition of (\ref{4.5}) according to (\ref{4.2}).
Note that  $q$, $p$ and $H$ are gauge invariant, while $\xi_{\K^\perp}$ matters only
up to conjugation by the elements of $K$, since
the gauge transformations generated by $\xi_\K$ act as
\be
(q,p,\xi_{\K^\perp}, L) \mapsto (q,p, e^\kappa \xi_{\K^\perp} e^{-\kappa}, e^\kappa L e^{-\kappa}),
\label{4.6}\ee
where $\kappa$ is an arbitrary $\K$-valued function.

The rational and trigonometric (hyperbolic)
$r$-matrices on the Cartan subalgebra of a simple
Lie algebra were classified in \cite{EV}, and  the corresponding
examples of spin Calogero models were described in \cite{LiXu,Li2}.
More recently, we studied \cite{FP} the family of systems based on
the $r$-matrices meeting the
compatibility and  non-degeneracy conditions.
In the quasi-triangular case all such  $r$-matrices
are provided (up to an irrelevant freedom in $\cR(q)\vert {\K}$) by the formula
\be
\cR(q)\vert {\K} =\frac{1}{2} \mathrm{id}_\K,\qquad
\cR(q)\vert {\K^\perp} = (1 - \theta^{-1} e^{-\ad_q} \vert \K^\perp)^{-1},
\label{4.7}\ee
where $\theta$ is an automorphism of $\G$ that preserves also the scalar product,
$\K$  lies  in the  fixpoint set of $\theta$, and the inverse that occurs is
well-defined for a non-empty open subset $\check \K \subset \K$.
In the general case this formula is due to Alekseev and Meinrenken \cite{AM},
its uniqueness property mentioned above was proved in \cite{FP}.

It turned out that
the spin Calogero models
associated with the $r$-matrices (\ref{4.7}) can be also
derived by Hamiltonian reduction of the geodesic system on
(an open submanifold of) $T^*G$.
The geodesics in question are simply the orbits of the one-parameter subgroups of $G$,
since the underlying metric on $G$ is induced from the invariant
bilinear form  $B_\G$ on $\G$ and is thus bi-invariant.
The reduction relies on the Hamiltonian  action of $G$
arising from twisted conjugations.
The twisted conjugation by $k\in G$ acts on the group manifold $G$ according to
\be
\Ad_k^\Theta: g\mapsto \Theta^{-1}(k) g k^{-1}\qquad
\forall g\in G,
\ee
if $\Theta\in \mathrm{Aut}(G)$ lifts $\theta\in \mathrm{Aut}(\G)$.
The Hamiltonian reduction method leads to a simple algorithm for integrating
the spin Calogero equation of motion with the aid of the geodesics of $G$.
The reader is referred to \cite{FP} for a detailed presentation of the
Hamiltonian reduction picture as well as for several examples.
The examples include systems built
on the non-trivial diagram automorphisms of the simply laced
simple Lie algebras and systems built on
the cyclic permutation automorphisms of semi-simple Lie algebras
composed of $N>1$ identical factors.
The former systems seem to be new, while the latter were studied
earlier in the $A_n$ case by Blom-Langmann \cite{BL} and by Polychronakos \cite{Poly} by means
of different methods.
It should be possible to quantize these systems with the aid of quantum Hamiltonian
reduction, which is one the topics of our interest for future work.

\bigskip
\bigskip
\noindent{\bf Acknowledgements.}
The work of L.F. was supported in part by the Hungarian
Scientific Research Fund (OTKA) under the grants
T043159, T049495,  M045596  and by the EU networks `EUCLID'
(contract number HPRN-CT-2002-00325) and `ENIGMA'
(contract number MRTN-CT-2004-5652).
B.G.P. is grateful for support by a CRM-Concordia Postdoctoral
Fellowship and he especially  wishes to thank J. Harnad for
hospitality in Montreal.

\end{document}